%% file: Yiran-main.tex
\documentclass[10pt,conference]{IEEEtran}

\usepackage[switch]{lineno}

\newif\ifanonymous
\anonymousfalse

\input{custom_config.sty}

\newcommand{\pp}{\textit{pp}}

\newcommand{\rqone}{\textit{RQ1: To what extent does runtime information improve LLM-based crash prediction and diagnosis in ML notebooks overall?}}
\newcommand{\rqtwo}{\textit{RQ2: How does the impact of runtime information on LLM-based crash prediction and diagnosis vary across different ML libraries and root causes?}}

\begin{document}

\ifanonymous
\linenumbers
\else
\fi

\sloppy
\title{CRANE-LLM: Runtime-Augmented LLMs for Crash Prediction and Diagnosis in ML Notebooks}

\ifanonymous
\author{{Anonymous Authors}}
\else
\author{\IEEEauthorblockN{Yiran Wang}
\IEEEauthorblockA{
\textit{Linköping University} \\ Linköping, Sweden \\
yiran.wang@liu.se}
\and
\IEEEauthorblockN{José Antonio Hernández López}
\IEEEauthorblockA{
\textit{University of Murcia} \\ Murcia, Spain \\
joseantonio.hernandez6@um.es}
\and
\IEEEauthorblockN{Ulf Nilsson}
\IEEEauthorblockA{
\textit{Linköping University} \\ Linköping, Sweden \\
ulf.nilsson@liu.se}
\and
\IEEEauthorblockN{Dániel Varró}
\IEEEauthorblockA{
\textit{Linköping University} \\ Linköping, Sweden \\
daniel.varro@liu.se}
}
\fi

\maketitle

\begin{abstract}
Jupyter notebooks have become popular for early machine learning (ML) development, enabling interactive and iterative experimentation. However, ML notebooks are prone to bugs, among which crashes are the most disruptive. Despite their practical importance, crash prediction and diagnosis in ML notebooks remain largely unexplored. We present CRANE-LLM, a runtime-augmented source code analysis approach that provides large language models (LLMs) with structured runtime information extracted from the notebook kernel, together with source code, to predict and diagnose crashes in a target cell before executing it. We evaluate CRANE-LLM on JunoBench, a benchmark of 111 Kaggle ML notebooks containing crashes across multiple ML libraries and crash types. Across three state-of-the-art LLMs (Gemini, Qwen, and GPT-5), our results show that runtime information significantly improves crash prediction and diagnosis performance by 7-10 percentage points in accuracy and 8-11 in F1-score, compared to using source code alone. The improvements are more pronounced when diagnosis is required, indicating that runtime context is especially valuable for reasoning about crash causes than merely predicting their presence.
\end{abstract}


\input{Main/Content/1_introduction}

\input{Main/Content/2_background}

\input{Main/Content/3_approach}

\input{Main/Content/4_experiments}

\input{Main/Content/5_results}

\input{Main/Content/6_discussion}

\input{Main/Content/7_relatedwork}

\input{Main/Content/8_conclusion}

\section*{Acknowledgment}
\ifanonymous
{Temporarily removed for double-blind review.}
\else
This work was partially supported by the Wallenberg AI, Autonomous Systems and Software Program (WASP) funded by the Knut and Alice Wallenberg Foundation, and the Software Center Project 61. The computations of open source LLMs were enabled by the Berzelius resource provided by the Knut and Alice Wallenberg Foundation at the National Academic Infrastructure for Supercomputing in Sweden (NAISS).
\fi

\bibliographystyle{IEEEtranDOI}
\bibliography{Main/references}

\end{document}

%% file: Main/Content/1_introduction.tex
\section{Introduction}

Machine learning (ML) techniques have become an integral part of modern software systems across domains such as medicine, marketing, and computer vision~\cite{Hanna25medicine, Kedi24media, ioannidou2017deep, pouyanfar2018survey}. 
However, ML programs are prone to bugs that arise from complex data processing, dynamic computation graphs, and rapidly evolving ML libraries~\cite{Morovati24bug, zhang2020machine}. 
Python-based ML programs, especially during the prototyping phase, are often developed in Jupyter Notebooks~\cite{Grotov22nbpy}, hereafter referred to as \textit{ML notebooks}. Notebook environments enable exploratory and incremental development through a cell-based execution model that maintains a persistent in-memory kernel state. However, this flexibility also introduces challenges for ensuring correctness. Since cells can be executed in any order, hidden dependencies and stateful objects often arise. 
In addition to these notebook-specific challenges, a recent empirical study~\cite{wang25mlcrash} shows that bugs in ML notebooks frequently stem from ML-specific issues rather than general Python bugs, such as data errors and ML library API constraint violations.

ML bugs vary in symptoms and severity. Some are silent bugs that degrade model performance, while others cause crashes that halt program execution and require immediate attention. Some crashes occur only after prolonged training~\cite{zhang20jobs}, making early detection or prediction crucial for efficient ML development. 
\emph{Early crash detection (crash prediction)}
is especially important in notebooks, where a crash in an executed cell may alter the notebook kernel to an inconsistent state, causing \emph{kernel state corruption}. 
Since there is no built-in support to save and restore the kernel state and reliable serialization is difficult, a crash has irreversible effects. Developers must restart the kernel and rerun \emph{all prior cells} (not just the crashing cell itself) to restore the correct state. Early crash detection avoids this by stopping execution before such kernel corruption occurs.
In this paper, we focus on \textit{crashes}, the most disruptive and frequently observed failure type~\cite{islamComprehensiveStudyDeep2019, Morovati24bug, desantanaBugAnalysisJupyter2024}, and aim to automatically predict them \emph{before} cell execution.

Several source code analysis approaches have been proposed to predict or detect bugs in ML programs, often targeting specific categories such as tensor shape mismatches~\cite{liu22shapetracer, jhooPyteaStaticAnalyzer2022, vermaShapeFlowDynamicShape2020, wuTensfaDetectingRepairing2021}, API misuses~\cite{bakerDetectFixVerify2022, liblit23left, wang21execution, weiDemystifyingDetectingMisuses2024}, neural network structural or training issues~\cite{Nikanjam21graph, manke25structurebugs, Wardat22deepdiagnosis, zhang21autotrainer, Braiek23fnn, Wardat21deeplocalize, cao22deepFD, wardat2023effectivedatadrivenapproachlocalizing, jahan2025improveddetectiondiagnosisfaults, sharmin25sa}. Work tailored specifically to notebooks has focused primarily on data leakages~\cite{suboticStaticAnalysisFramework2022, suboticStaticallyDetectingData2022, yang23leakagenb, Drobnjakovic24leakagenb}, execution consistency~\cite{macke21finegrain, liblit23left}, reproducibility and dependency management~\cite{zhuRestoringExecutabilityJupyter2021, wang21execution}, or code quality and maintainability~\cite{quarantaPynblintQualityAssurance2024, patra22nalin}. However, no existing work has addressed crash prediction in notebooks for a diverse set of ML libraries and bug types.

Recent evidence suggests that traditional static analysis is often inadequate for predicting ML bugs that depend on runtime data~\cite{weiDemystifyingDetectingMisuses2024}, while dynamic approaches cannot predict and prevent ML crashes before execution. 
Notebook environments inherently provide \emph{rich runtime information} in the kernel state after executing a few cells, offering the opportunity to predict and prevent crashes before subsequent cells are run. 
A new idea paper~\cite{wang24runtime} highlighted this potential, but it offers only some initial insights, focusing solely on predicting \emph{TensorFlow}-related shape bugs, limiting its generality and practicality.

Prior studies~\cite{Kochhar16FLrational, Kang24FLexplain} emphasize that bug prediction and localization are often insufficient as developers also need rationales why a particular bug occurs and why the localized code is faulty to effectively understand and fix it. 
However, few existing techniques provide such level of diagnosis.
The empirical study by Jahan~\etal~\cite{Jahan25empiricalbuglocalDL} highlights that large language models (LLMs) have potential in bug diagnosis.

This paper introduces \emph{CRANE-LLM, an LLM-based, runtime-augmented crash prediction and diagnosis framework} for ML notebooks.
CRANE-LLM aims to predict crashes in a target code cell \textit{before} its execution by using LLMs to analyze the  source code of notebooks together with runtime information (e.g., data attributes) extracted from the notebook kernel, enabling context-aware crash prediction and diagnosis. 
The end-to-end latency of CRANE-LLM typically takes only a few seconds per notebook (e.g., 1.6s on average using GPT-5).
This enables proactive error prediction before cell execution, preventing kernel state corruption, and avoiding unnecessary waiting time, such as when a crash occurs after a prolonged model training process.
In summary, this paper makes the following contributions:
\begin{itemize}[leftmargin=*]
\item We propose CRANE-LLM, a runtime-augmented LLM framework that enables pre-execution crash prediction and diagnosis for ML notebooks.
\item We evaluate the impact of runtime information on three state-of-the-art LLMs using JunoBench~\cite{wang2025junobenchbenchmarkdatasetcrashes}, a benchmark of 111 real ML notebooks with reproducible crashes.
\item We analyze the effectiveness of runtime information across different ML libraries and crash root causes, demonstrating its value for improving LLM-based crash reasoning.
\end{itemize}

\textbf{Data availability.}
To ensure reproducibility, we released our replication package on GitHub~\cite{wang2025code}.

%% file: Main/Content/2_background.tex
\section{Background}


\input{Main/Sub-content/fig_notebook_example}

\paragraph*{ML notebooks}
A notebook consists of a sequence of \emph{cells} that contain either \emph{executable code} or \emph{documentation}. 
Code cells can be executed independently or sequentially within a shared kernel session. The execution order is indicated by the \emph{execution count} displayed beside each executed cell. 
Executing a cell can produce \emph{outputs} such as textual prints, visualizations, or error messages when the cell crashes. 
\autoref{fig:notebook_example} (left) shows a real notebook from JunoBench~\cite{wang2025junobenchbenchmarkdatasetcrashes}. In this notebook, the developer constructs a neural network using \emph{TensorFlow/Keras}. The cells with execution counts have been executed. First, an image dataset is loaded, split into training and test sets, and data generators are defined for preprocessing and augmentation. Then a training iterator \lstinline[style=mystylecode]{train_images} is created from the training split, followed by the definition of a neural network model. The model is configured with five output units and compiled using the categorical cross-entropy loss function. 
The last cell that has not been executed yet attempts to train the model by invoking \lstinline[style=mystylecode]{model.fit} with previously defined training iterator \lstinline[style=mystylecode]{train_images} and validation data.

\paragraph*{Runtime information in ML notebooks}
Unlike Python scripts that execute linearly from start to finish, notebooks allow nonlinear and selective executions. The kernel maintains a persistent execution state across cells~\cite{macke21finegrain, wang24runtime}.
This kernel state represents a complete snapshot of the program's execution at a given time and can be accessed while the session is active.
However, directly inspecting kernel variables is often impractical, as it produces either overly verbose or insufficiently informative outputs. 
As shown in~\autoref{fig:notebook_example} (middle), some objects (e.g., dataframes) display all their values when printed, whereas others (e.g., \emph{Keras} iterators) reveal only minimal type information. 
While functions like \lstinline[style=mystylecode]{vars()} expose internal attributes, they may produce excessive low-level values. 
This makes it challenging to obtain a meaningful and appropriately detailed summary of the kernel state for downstream analysis. 

To address this, we \emph{extract runtime information} to selectively summarize object properties into structured, concise, and interpretable representations. 
This extracted runtime information enables crash prediction before cell execution.
For example, in~\autoref{fig:notebook_example} (right), the model output layer defines five neurons, but the number of classes in \lstinline[style=mystylecode]{train_images} cannot be inferred from the source code. Predicting whether the last cell will crash therefore requires extracting generator attributes \lstinline[style=mystylecode]{num_classes} from the kernel state.

\input{Main/Sub-content/fig_notebook_example2}

\paragraph*{Kernel state corruption in notebooks} \label{sec_background: kernel_corruption}
In notebooks, executing a crashing cell can leave the kernel in a partially updated or inconsistent state, as successfully executed statements (before the crashing line) may have already mutated global variables or model parameters, while subsequent statements are skipped. 
While direct measurements of kernel corruption frequency in real-world development are scarce, a prior study~\cite{freund2026flowbook} reports that the vast majority ($\sim74\%$) of real-world notebooks contain cells that can lead to corrupted kernel state.

\autoref{fig: notebook_example2} shows an example of kernel state corruption. The notebook first loads the data and constructs a \emph{PyTorch} neural network. Inspecting the model parameters after executing the first two cells shows their initial state (left side of the figure). 
When the next cell is executed, 
the \lstinline[style=mystylecode]{optimizer.step()} call run without error and updates the model parameters, 
but the cell crashes during validation due to incompatible input dimensions. 
Inspecting the model parameters again (right side of the figure) shows that they have changed,
leaving the kernel in an inconsistent state.
If the developer continues to fix the bug and re-execute only the crashing cell, the optimizer updates the model parameters again. Because the first iteration partially executed before crashing, the parameters now reflect a partial update followed by a repeated update, which can silently corrupt the model state and produce incorrect or unpredictable behavior.

Because Jupyter provides no built-in mechanism to snapshot, roll back, or reliably serialize the full kernel state, such corruption is irreversible. As a result, restoring correctness requires restarting the kernel and re-executing all prior cells, including potentially expensive preprocessing steps. This makes crashes particularly costly in notebook-based ML workflows and motivates techniques for crash prediction before cell execution to prevent kernel state corruption.

%% file: Main/Sub-content/fig_notebook_example.tex
\begin{figure*}
    \centering
    \includegraphics[width=\textwidth]{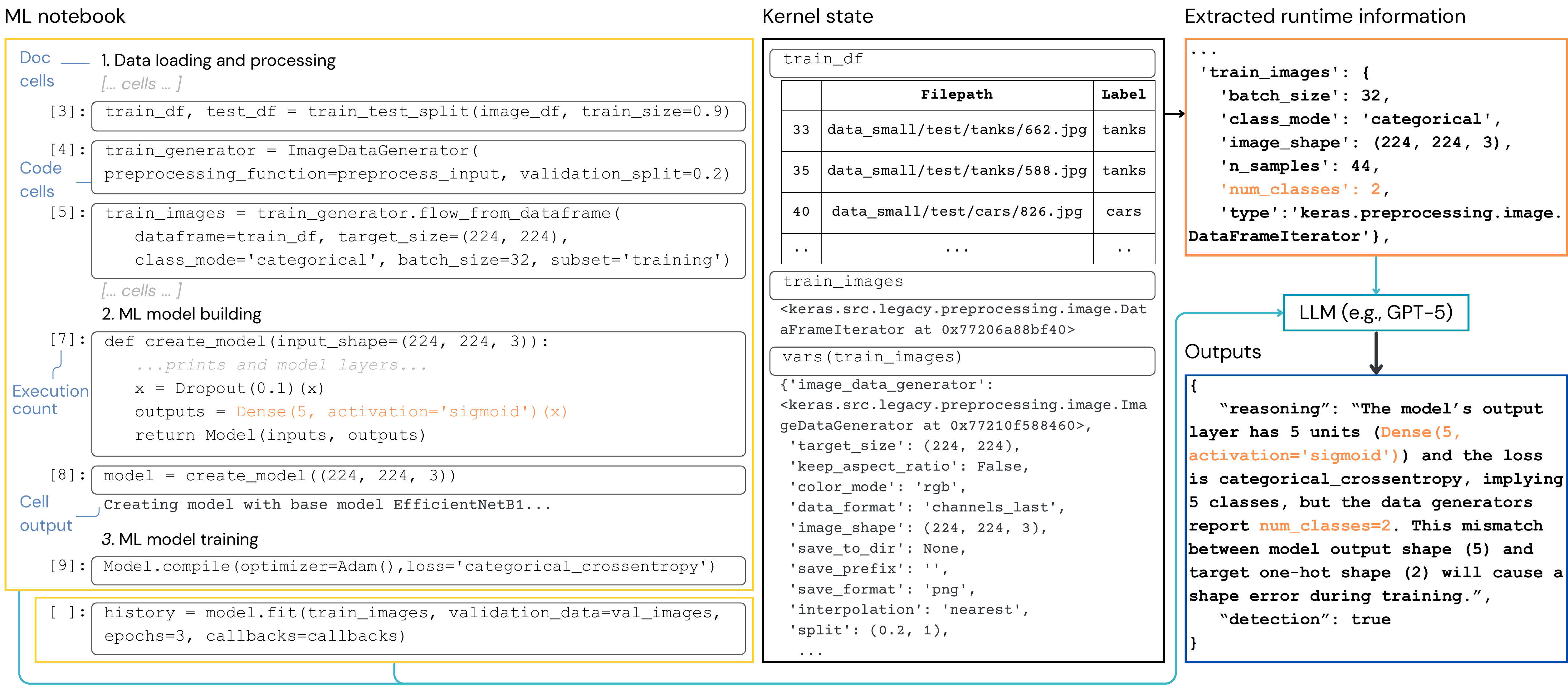} 
    \caption{A JunoBench notebook example showing the notebook structure (left), kernel state (middle), and showcase the CRANE-LLM pipeline.}
    \label{fig:notebook_example}
\end{figure*}

%% file: Main/Sub-content/fig_notebook_example2.tex
\begin{figure}
    \centering
    \includegraphics[width=\linewidth]{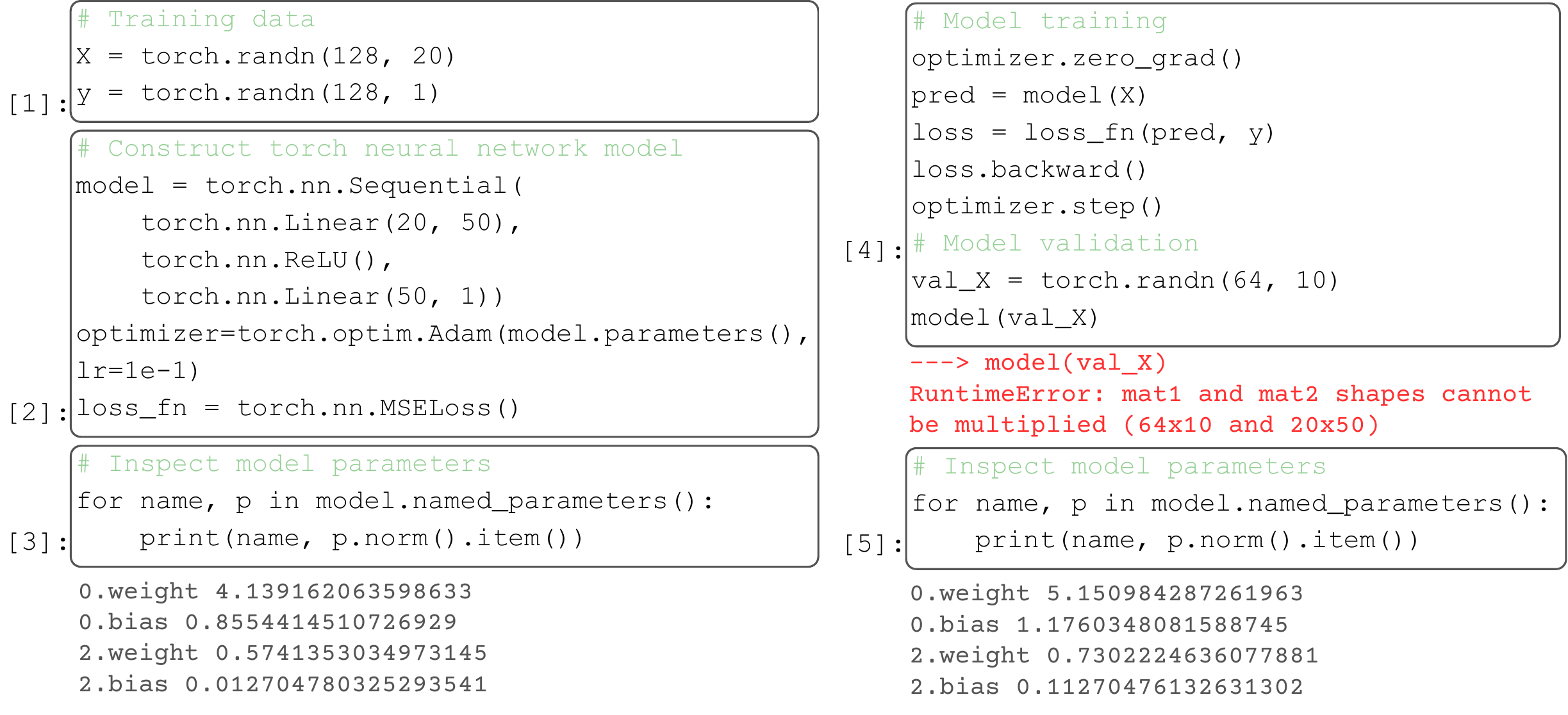} 
    \caption{A simulated notebook example showing how a crashing cell can irreversibly corrupt the kernel state by updating model parameters.}
    \label{fig: notebook_example2}
\end{figure}

%% file: Main/Content/3_approach.tex
\section{The CRANE-LLM Approach}

\input{Main/Sub-content/fig_CRANE_overview}

\autoref{fig: approach_overview} shows the overview of CRANE-LLM. Given a set of successfully executed code cells and a target cell as input, CRANE-LLM attempts to predict whether executing the target cell would crash. If a crash is predicted, it provides a diagnosis that allows developers to inspect the potential flaw and, if in agreement, remediate the bug before executing the cell. CRANE-LLM consists of two main steps: \emph{runtime information extraction} (\autoref{sec:runtime-info-extract}) and LLM-based \emph{crash prediction and diagnosis} (\autoref{sec:crash-detect-diag}). 


\subsection{Runtime Information Extraction}
\label{sec:runtime-info-extract}
First, we extract runtime information from the current Jupyter notebook kernel state (i.e., the program state generated by executed code cells). 
Because the full kernel state can include all objects and their nested data structures, it is often prohibitively large to feed into an LLM. Providing the entire context would not only increase token usage and computational or economic cost, but also introduce noise that could distract the model from identifying information relevant to the task. 

We first filter the kernel state to retain only information relevant to the target cell by parsing its abstract syntax tree (AST) to identify referenced variables, function calls, and attributes, and collecting their runtime data. 
As shown in~\autoref{fig:notebook_example}, raw runtime inspection can reveal either excessively details or almost nothing, so we summarize relevant objects into structured abstractions.

Our choice of target object types for runtime summarization is motivated by a
prior study~\cite{wang25mlcrash}, which identifies data-related bugs, such as tensor shape mismatches and dataset misalignments, are among the most frequent, and \emph{TensorFlow/Keras}, \emph{PyTorch}, and \emph{Scikit-learn} as the three most challenging ML libraries. 
Accordingly, we extract runtime information for common ML data structures, including tensors (\emph{PyTorch}, \emph{TensorFlow/Keras}), arrays (\emph{NumPy}), and dataframes (\emph{Pandas}), as well as ML-specific objects that directly affect execution outcomes, such as image data iterators and datasets (\emph{TensorFlow/Keras}), iterable datasets and data loaders (\emph{PyTorch}), and label encoders and models (\emph{Scikit-learn}). 

\paragraph{Tensors, arrays, and dataframes}
For tensors and arrays of arbitrary dimensionality, including \lstinline[style=mystylecode]{numpy.ndarray}, \lstinline[style=mystylecode]{torch.Tensor}, \lstinline[style=mystylecode]{tensorflow.Tensor}, and \lstinline[style=mystylecode]{pandas.Series}, we record their shapes, \lstinline[style=mystylecode]{dtypes} (data type of each element, e.g., \lstinline[style=mystylecode]{float32} or \lstinline[style=mystylecode]{bool}), and presence of missing values. 
For one-dimensional arrays, we additionally characterize value types (binary, categorical, or continuous), unique value counts (for binary or categorical value type), and value ranges (for continuous value type).
For two-dimensional data structures such as \lstinline[style=mystylecode]{pandas.DataFrame}, we summarize each column's value type, unique value counts or numeric range, and representative example values.
To control prompt length to LLMs, we set a configurable upper limit on the number of columns included per dataframe (20 by default).

\paragraph{ML library-specific data}
For \emph{TensorFlow/Keras}, we summarize \lstinline[style=mystylecode]{ImageDataGenerator} iterators (such as \lstinline[style=mystylecode]{DirectoryIterator} and \lstinline[style=mystylecode]{DataFrameIterator}) by recording their number of samples, number of classes, batch size, image shape, and target size.
We also summarize \lstinline[style=mystylecode]{tensorflow.data.Dataset}. 
Since actual values are only accessible via iteration (e.g., \lstinline[style=mystylecode]{take()}), to avoid consuming data, we only non-intrusively extract metadata such as \lstinline[style=mystylecode]{element_spec}, which provides tensor shape, \lstinline[style=mystylecode]{dtype}, and \lstinline[style=mystylecode]{TensorSpec} attributes.

For \emph{PyTorch}, we distinguish between map-style datasets and iterable datasets. Map-style datasets inherit from \lstinline[style=mystylecode]{torch.utils.data.Dataset} and implement both \lstinline[style=mystylecode]{__getitem__} and \lstinline[style=mystylecode]{__len__}, allowing safe random access to individual samples. For these datasets, we record the dataset size using \lstinline[style=mystylecode]{len(dataset)} and probe a configurable small number of samples (e.g., 10) via \lstinline[style=mystylecode]{dataset[i]} to summarize input, label, and shape. We optionally simulate batching using \lstinline[style=mystylecode]{collate_fn} to capture batch-level shapes. 
For iterable datasets such as \lstinline[style=mystylecode]{torch.utils.data.IterableDataset} or data loaders (e.g., \lstinline[style=mystylecode]{torch.utils.data.DataLoader}), accessing the samples consumes those elements. Therefore, we only record static metadata such as class names and schema without materializing actual data.

For \emph{Scikit-learn}, we capture metadata from objects such as \lstinline[style=mystylecode]{LabelEncoder} (e.g., number of classes and their \lstinline[style=mystylecode]{dtype}) and models (e.g., class name, fit status, and fitted attributes such as number of input features and output classes). 


\textbf{Example.}
\autoref{fig:notebook_example} (top-right) shows an example of extracted runtime information.
In the notebook (left), we aim to predict crashes in the last code cell (i.e., \textit{target cell}). We exact relevant runtime information from the kernel state (middle).
For instance, \lstinline[style=mystylecode]{train_images} (\lstinline[style=mystylecode]{DataFrameIterator}) relevant to the target cell is summarized with attributes such as batch size, image shape, number of samples, and number of classes. 

\subsection{Crash Prediction and Diagnosis} 
\label{sec:crash-detect-diag}
After extracting runtime information, we construct a prompt that combines three key elements: (1) the ordered sequence of previously executed cells, (2) the target cell under prediction, and (3) extracted runtime information. This prompt is used to query an LLM to predict whether the target cell will crash.

Designing the prompt for CRANE-LLM is guided by two design considerations. First, providing the LLM with \emph{successfully executed cells in execution order} allows it to analyze the source code and reason about dependencies that may influence the target cell~\cite{bubeck2023sparks}. This reflects the incremental execution model of notebooks and provides the necessary code context to understand data flow and variable dependencies. Second, the prompt includes \emph{relevant runtime information}, such as object types and data attributes, to expose execution states that are typically unavailable from source code alone. This information is particularly important for ML notebook crashes that depend on concrete runtime properties, such as tensor shape mismatches or violations of data constraints.

The prompt explicitly instructs the LLM to \emph{reason step-by-step} to encourage chain-of-thought reasoning, which prior work has shown improves performance on complex tasks~\cite{Wei22CoT}. In ML crash prediction and diagnosis, this helps analyze issues arising from interactions between data transformations, training, and model expectations. For example, a shape mismatch occurs when an input tensor reaches a layer expecting a different shape, or when the number of classes produced by a data generator does not match a model’s output layer. 

For LLMs that support structured output through APIs (e.g., Gemini), we enforce a predefined JSON schema in the API call. For other LLMs, we prompt the LLM to response in the same JSON format. The LLM output consists of a concise diagnostic explanation (i.e., \emph{reasoning}) followed by a binary crash \emph{prediction} verdict (\textit{true} only when the model is confident that the target cell will crash, and \textit{false} otherwise). This standardized format ensures consistency across models and improves the interpretability of LLM outputs. 
Detailed prompts can be found in our GitHub repository~\cite{wang2025code}.

\textbf{Example.}
In~\autoref{fig:notebook_example},
we provide the executed cells, extracted runtime information, and the target cell together as input to an LLM such as GPT-5.
The LLM predicts that the target cell will crash. It diagnoses that the output layer of the network produces five logits, while the training data iterator reports only two classes. This mismatch leads to an incompatibility when computing the loss function. GPT-5 highlights that the runtime information, in particular \lstinline[style=mystylecode]{num_classes=2} of the iterator, is decisive for the prediction.

This example illustrates that while the source code alone is insufficient to reveal inconsistencies related to data, the extracted runtime information confirms the mismatch with certainty, enabling the LLM to predict the crash correctly.

%% file: Main/Sub-content/fig_CRANE_overview.tex
\begin{figure}[t]
    \centering
    \includegraphics[width=\columnwidth]{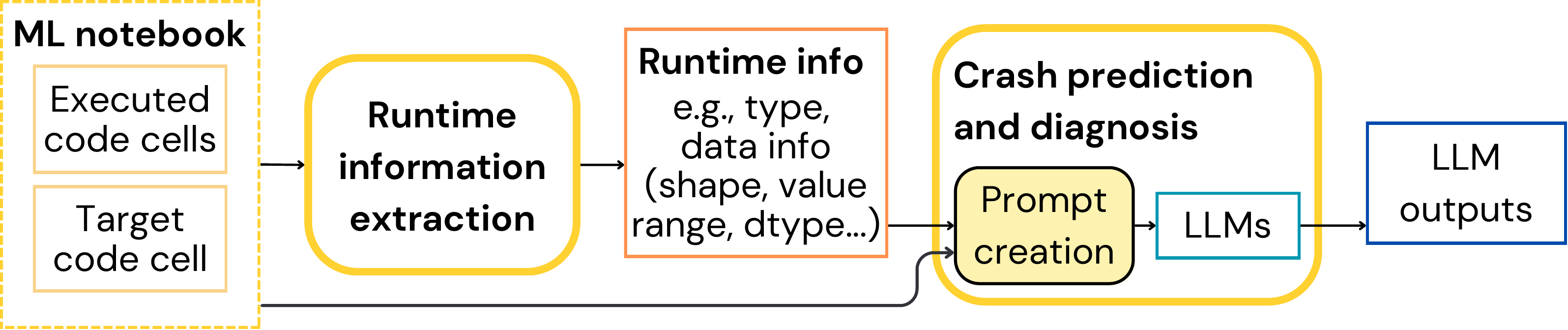} 
    \caption{Overview of CRANE-LLM.}
    \label{fig: approach_overview}
\end{figure}

%% file: Main/Content/4_experiments.tex
\section{Evaluation Procedure}
Next, we present the research questions, dataset, LLMs, evaluation protocol, and experimental procedure.

\subsection{Research Questions (RQs)}
\label{subsec: rqs}
To assess the performance of CRANE-LLM, our experiments aim to address the following research questions:

\rqone
    
    \underline{Motivation:} Many crashes in ML notebooks arise from mismatches between code assumptions and concrete runtime states, such as tensor shapes or data dimensionality. However, it is unclear whether providing explicit runtime information yields benefits beyond what LLMs can already approximate from static code alone, or whether the added complexity of extracting runtime context is justified. By systematically quantifying the impact of runtime information on crash prediction and diagnosis performance, we assess when and to what extent runtime-aware LLMs offer practical advantages.

\rqtwo

    \underline{Motivation:} ML notebook crashes arise from diverse root causes, such as data confusion errors and API misuse, and occur while using a variety of ML libraries with distinct abstractions and execution models. This suggests that runtime information may not benefit all crashes equally. Therefore, we investigate how the impact of runtime information varies across ML libraries and root causes, shedding light on which crash patterns benefit most from runtime context and whether runtime-aware approaches should be specialized or generalized.

\subsection{Dataset}
\label{subsec: dataset}
To evaluate CRANE-LLM, we use JunoBench~\cite{wang2025junobenchbenchmarkdatasetcrashes}, a benchmark dataset of real-world Kaggle ML notebooks containing 111 buggy notebooks, each with exactly one crashing code cell, along with their corresponding fixed versions, resulting in 222 notebooks in total. To the best of our knowledge, at the time being, JunoBench is the only publicly available benchmark specifically designed to study crashes in ML notebooks.

We adopt JunoBench for the following reasons: 
(1) It is curated from Kaggle notebooks in their original execution form, matching our target use case. 
(2) It balances crashes across major ML libraries, including 15 cases each from \emph{TensorFlow/Keras}, \emph{PyTorch}, \emph{Scikit-learn}, \emph{NumPy} and \emph{Pandas}; 16 cases collectively from \emph{other libraries} (\emph{Seaborn}, \emph{Matplotlib}, \emph{Statsmodels}, \emph{TorchVision}, and \emph{LightGBM}); and 20 cases from \emph{NBspecific} (notebook out-of-order execution). 
(3) Each buggy notebook is accompanied by a sequence of ordered executed code cells that reliably reproduce the crash, along with a verifiable fix. All instances are executable within a unified Docker environment, ensuring consistency and facilitating practical evaluation. 
(4) JunoBench provides empirical annotations, including \emph{crash type}, \emph{root cause}, the \emph{ML library} used at the crash point, as well as ground truth labels for crash existence (\textit{true} for buggy and \textit{false} for fixed notebooks) and crash diagnoses explaining why each crash occurs. 
These labels enable performance evaluation of both crash existence prediction and crash diagnosis.

\subsection{LLMs in CRANE-LLM}
\label{subsec: LLMs}
We evaluated CRANE-LLM using three LLMs:

 \begin{itemize} 
     \item \textbf{Gemini-2.5-Flash}~\cite{comanici2025gemini}
     , a commercial model from Google.
     \item \textbf{GPT-5}~\cite{openai2025gpt5}
     , a state-of-the-art OpenAI model.
     \item \textbf{Qwen-2.5-Coder-32B-Instruct}~\cite{qwen2025qwen25}
     , a leading \textit{open-source} instruction-tuned model from Alibaba Cloud.
 \end{itemize}

These LLMs represent a broad spectrum: a high-throughput commercial model optimized for efficiency (Gemini), a widely adopted industry-standard model with extensive capabilities (GPT-5), and an open-source alternative facilitating reproducibility and transparency (Qwen). This selection allows us to assess whether CRANE-LLM generalizes across models with different trade-offs in speed, capacity, and accessibility.

LLM APIs may produce different outputs for identical inputs due to inference-time nondeterminism. To improve robustness, we query each notebook \emph{five times} and use \emph{majority voting} over the predictions. For Gemini and Qwen, we set the temperature to 0.01 to reduce stochastic variation, while GPT-5 does not expose temperature control. We apply the same majority-vote procedure to all LLMs for a fair comparison.

\subsection{Evaluation}
\label{subsec: evaluation}
We evaluate the crash prediction and diagnosis performance of CRANE-LLM under two settings: 
(1) \emph{crash prediction} (i.e., crash existence prediction) alone, and (2) \emph{end-to-end crash prediction and diagnosis} together.

Following prior work on bug prediction and localization~\cite{yang24mulbugs, li21faulloc, li19deepfl}, we use metrics including precision, recall, F1-score, and accuracy. 
Let TP, FP, TN, and FN denote true positives, false positives, true negatives, and false negatives, these metrics are defined as follows:
\begin{equation*}
    \text{prec} = \frac{\text{TP}}{\text{TP}+\text{FP}}\quad\text{recall} = \frac{\text{TP}}{\text{TP}+\text{FN}} \quad F_1 = \frac{2 \cdot \text{prec} \cdot \text{recall}}{\text{prec} + \text{recall}}
\end{equation*}
$$\text{accuracy} = \frac{\text{TP} + \text{TN}}{\text{TP} + \text{TN} + \text{FP} + \text{FN}}$$

\subsubsection{Crash prediction evaluation}
\label{subsubsection: evaluation_detection}
Each LLM output is a JSON object containing: (i) a boolean field \emph{prediction} (i.e., \textit{true} or \textit{false}) indicating whether a crash is predicted, and (ii) a free-text field \emph{diagnosis} describing why the crash occurs.

Crash prediction performance is evaluated automatically by comparing only the \emph{prediction} field against the ground truth crash existence label provided by JunoBench (see~\autoref{subsec: dataset}). This fully automated procedure enables exact, scalable, and reproducible evaluation of crash predictions across all outputs. \emph{Diagnosis} field does not affect crash prediction metrics.
Specifically, each crash prediction output is classified into the following two outcomes:

\begin{itemize}
    \item \textbf{Correct:} The crash prediction is correct.
    \item \textbf{Wrong:} The crash prediction is incorrect.
\end{itemize}

For metric calculation under crash prediction evaluation:

\begin{itemize}
    \item TP: a crash is predicted; ground truth indicates a crash.
    \item FP: a crash is predicted; ground truth shows no crash.
    \item TN: no crash is predicted; ground truth shows no crash.
    \item FN: no crash is predicted; ground truth indicates a crash.
\end{itemize}

\subsubsection{End-to-end (crash prediction and diagnosis) evaluation}
End-to-end evaluation assesses whether the LLM output both (1) correctly predicts a crash and (2) provides a correct diagnosis when a crash is correctly predicted.
Each output is classified into one of the following outcomes: 

\begin{itemize}
    \item \textbf{Correct:} The crash prediction is correct and, when prediction is \textit{true}, the diagnosis output is fully aligned with ground truth diagnosis in JunoBench.
    \item \textbf{Partially correct:} The crash prediction is correct, and when it is \emph{true}, the diagnosis output is partially correct but contains incorrect elements.
    \item \textbf{Reasoning wrong:} The crash prediction is correct, but the diagnosis output is entirely inaccurate or unrelated.
    \item \textbf{Wrong:} The crash prediction is incorrect, regardless of the diagnosis output correctness.
\end{itemize}

We treat both \emph{partially correct} and \emph{reasoning wrong} outcomes as \emph{wrong} in metric calculation, since incorrect or irrelevant diagnosis could mislead developers during debugging. However, separately tracking them provides additional insight into the reasoning tendencies of the LLMs. Therefore, for metric computation, we reduce these outcomes to a binary decision reflecting practical usefulness:

\begin{itemize}
    \item TP: a crash is correctly predicted and \emph{correctly diagnosed}.
    \item FP: a crash is predicted; ground truth indicates no crash.
    \item TN: no crash is predicted; ground truth indicates no crash.
    \item FN: a crash exists but is not predicted or \emph{misdiagnosed}.
\end{itemize}

End-to-end evaluation proceeds in two stages. First, crash prediction evaluation is determined automatically as described in~\autoref{subsubsection: evaluation_detection}. Second, for the subset of cases where both the prediction and ground truth indicate a crash, crash diagnosis evaluation is required, which is evaluated by human annotators.

\subsubsection{Human evaluation of crash diagnosis}
\label{subsubsection: evaluation_disgnosis}
Crash diagnosis evaluation assesses the quality of the explanatory reasoning generated by the LLMs and therefore requires human semantic judgment. The diagnosis is evaluated for the correctly predicted crash cases. Each such LLM output is classified into one of three categories: \emph{correct}, \emph{partially correct}, or \emph{reasoning wrong}, according to its consistency with the ground-truth diagnosis labels provided in JunoBench (see~\autoref{subsec: dataset}).

We employ two human evaluators (both are authors of this paper). 
First, all applicable predictions are evaluated by one evaluator. To assess evaluation consistency, a second evaluator independently evaluates a statistically representative subset of size $n$ drawn from the full set of $N$ diagnoses requiring evaluation. The target sample size $n$ is estimated using standard statistical sampling with a 95\% confidence level and a 5\% margin of error.
Finally, inter-annotator agreement is measured using Cohen’s $\kappa$ and reported to quantify evaluation reliability.

\input{Main/Sub-content/fig_experiment}

\subsection{Experiment Procedure}
\label{subsec: exp_procedure}
An overview of the experiment procedure is presented in~\autoref{fig: exp_overview}. 
CRANE-LLM performs the crash prediction and diagnosis task for all 222 notebooks (both buggy and fixed versions) in JunoBench. The evaluation considers both the \emph{end-to-end crash prediction and diagnosis} and the \emph{crash prediction-only} settings.
We compare LLM performance with (+RT) and without (-RT) access to runtime context. 

\paragraph{RQ1: Runtime info impact on crash prediction and diagnosis}
We compare CRANE-LLM, the \emph{LLM (+RT)} configuration, against a \emph{baseline} which does not use runtime information, referred to as \emph{LLM (-RT)}.
The baseline uses the same CRANE-LLM pipeline but includes only the executed code cells, without any extracted runtime information.
Using executed cells ensures that the input reflects the actual code context that leads to the crash or the applied fix, since including non-executed or irrelevant cells could introduce additional errors and cause the fixed version to fail. This setup isolates the effect of extracted runtime information and allows us to measure the extent to which it improves crash prediction and diagnosis performance of LLMs for ML notebooks.

Overall, the experimental setup for RQ1 is:
\begin{itemize}
    \item LLM (+RT): CRANE-LLM, with runtime information
    \item LLM (-RT): baseline, without runtime information
\end{itemize}

Outputs from each LLM are generated for every notebook in both experimental setups, resulting in 1,332 \emph{cases} ($222$ notebooks $\times$ $2$ setups $\times$ $3$ LLMs). Each case consists of five independently generated \emph{instances}. For each case, the final outcome is determined using a majority-vote strategy across its five instances.

Crash prediction performance is evaluated using the automatic procedure described in~\autoref{subsubsection: evaluation_detection}. For end-to-end crash prediction and diagnosis evaluation, we additionally conduct manual diagnosis assessment for cases that correspond to buggy notebooks and are correctly predicted as crashes. This results in a total of 501 cases. Manual diagnosis evaluation follows the protocol described in \autoref{subsubsection: evaluation_disgnosis}.

Specifically, one human evaluator assessed all 501 cases. To assess evaluation consistency, we computed a statistically representative sample of 218 cases and applied stratified sampling to ensure balanced representation across three LLMs and two experimental setups. Fractional sample counts were rounded to integer values, resulting in a final sample size of 222 cases. All sampling procedures were conducted using fixed random seeds to ensure reproducibility.~\autoref{tb: sampling} reports the detailed stratified sampling distribution.
Then the second human evaluator independently evaluated the 222 cases.
Inter-rater agreement was measured using Cohen’s kappa coefficient, yielding $\kappa = 0.89$, which indicates \emph{almost perfect} agreement under standard interpretation guidelines.

\input{Main/Sub-content/tb_sampling}
\paragraph{RQ2: Runtime info impact on ML libraries and root causes}
RQ2 is a post-hoc analysis based on the outputs obtained in RQ1. Using the same crash prediction and diagnosis results, we stratify performance by the \emph{ML library in use} and \emph{root cause} labels provided in JunoBench. This analysis examines how the effectiveness of runtime information varies across different ML libraries and root causes.

%% file: Main/Sub-content/fig_experiment.tex
\begin{figure}
    \centering
    \includegraphics[width=\columnwidth]{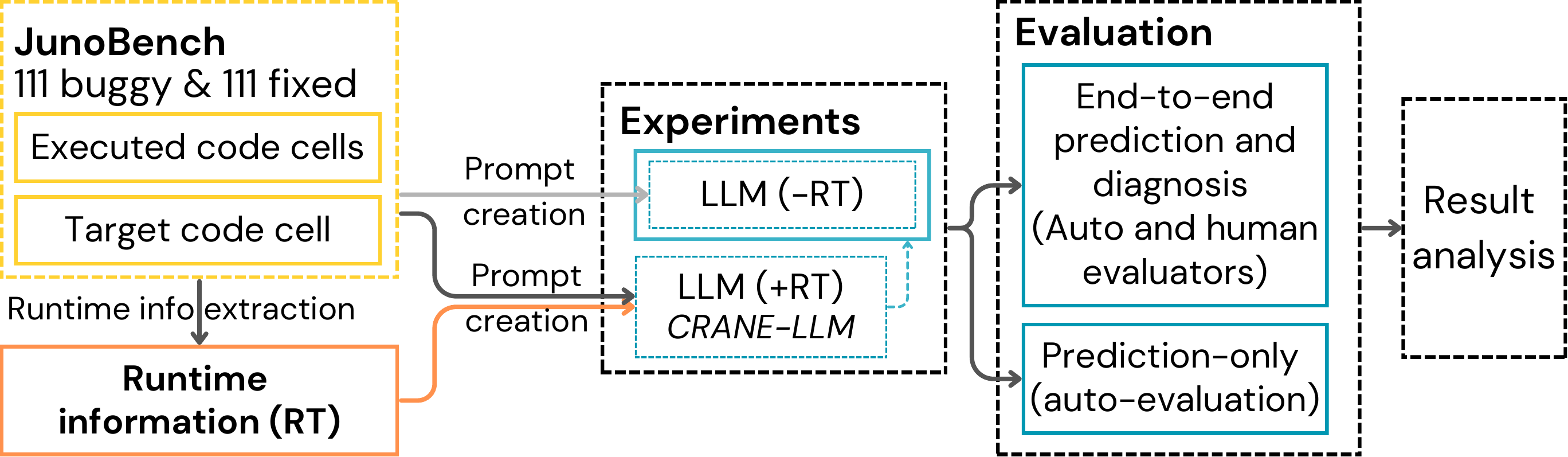} 
    \caption{Overview of experimental setup.}
    \label{fig: exp_overview}
\end{figure}

%% file: Main/Sub-content/tb_sampling.tex
\begin{table}[t]
\small
\centering
\caption{Stratified sampling distribution for human evaluation.}
\label{tb: sampling}
\begin{tabular}{c|c|c|c}
    \toprule
    \thead{LLM-Setting} & \thead{\#Applicable} & \thead{Proportion} & \thead{\#Sampled} \\
    \midrule
    Gemini (-RT)      & 90 & 0.180 & 40 \\\hline
    Gemini (+RT)        & 99 & 0.198 & 44 \\\hline
    Qwen (-RT)        & 71 & 0.142 & 31 \\\hline
    Qwen (+RT)          & 74 & 0.148 & 33 \\\hline
    GPT-5 (-RT)           & 79 & 0.158 & 35 \\\hline
    GPT-5 (+RT)            & 88 & 0.176 & 39 \\
    \midrule
    \textbf{Total}              & \textbf{501} & \textbf{1.000} & \textbf{222} \\
    \bottomrule
\end{tabular}
\end{table}

%% file: Main/Content/5_results.tex
\section{Results}

This section presents the results for each research question and discusses the threats to validity of the study.

\subsection{RQ1: Runtime info impact on crash prediction and diagnosis}
\label{subsec: results_rq1}

\input{Main/Sub-content/tb_results_overall}

As shown in~\autoref{tb: results_overall}(a), augmenting LLMs with runtime information (\textit{+RT}) consistently improves performance in {\textit{crash prediction and diagnosis}} in ML notebooks for all evaluated LLMs for both accuracy and F1-score. 
To assess statistical significance, we performed paired comparisons between the \textit{+RT} and \textit{-RT} settings using the two-sided exact McNemar test on the LLM outcomes, following prior work~\cite{abhi2025mcnemar, steenhoek2024mcnemar}. The improvements gained by runtime information for all three LLMs are significant ($p<0.05$).

The largest gains are observed for Qwen, with accuracy increasing by 9.4 percentage points (\pp) and F1-score by 9.3\pp. GPT-5 and Gemini also benefit substantially, achieving accuracy gains of 7.7\pp~and 7.2\pp, and F1-score gains of 10.6\pp~and 8.1\pp, respectively.
Across all three LLMs, runtime information improves both precision and recall, though their relative gains differ. For Gemini and GPT-5, the improvement is more pronounced in recall (+10.8\pp~and +14.4\pp) than in precision (+5.8\pp~and +4.8\pp), indicating a higher rate of correctly identifying crashing notebooks when runtime information is available. In contrast, Qwen exhibits larger gains in precision than recall (+16.3\pp~vs. +5.4\pp), reflecting a greater reduction in false positives. These results suggest that runtime information affects different LLMs in distinct ways, while consistently improving overall effectiveness.

\input{Main/Sub-content/tb_results_diagnosis_subcat}

In addition to metric-level improvements, our further analysis (\autoref{tb: results_diagnosis_subcat}) shows that runtime information reduces two categories of incorrect diagnosis predictions including \emph{partially correct} and \emph{reasoning wrong}, across all evaluated LLMs, with one case remaining unchanged. Both categories correspond to outputs where the crash prediction is correct but the crash diagnosis is flawed. Their simultaneous reduction indicates that runtime information contributes not only to improved prediction accuracy, but also to more reliable diagnostic reasoning.

\important{RQ1/1}{Runtime information improves LLM-based crash prediction and diagnosis across all evaluated LLMs. Accuracy increases by 7.2\pp, 9.4\pp, and 7.7\pp, and F1-score by 8.1\pp, 9.3\pp, and 10.6\pp~for Gemini, Qwen, and GPT-5, respectively. Gemini and GPT-5 benefit primarily from improved recall, while Qwen benefits more from increased precision.}

\autoref{tb: results_overall}(b) shows the crash prediction-only results, where the task is to predict whether a notebook crashes, without requiring diagnosis. In this setting, runtime information yields smaller gains (+4 to +8\pp~in accuracy and +5 to +7\pp~in F1-score) compared to the joint task of crash existence prediction and diagnosis (+7 to +10\pp~in accuracy and +8 to +11\pp~in F1-score). 
The McNemar test indicates that the improvements obtained by Qwen are statistically significant, whereas those for Gemini and GPT-5 are not. 
This gap suggests that runtime information is more beneficial when models are required to reason about underlying crash causes, rather than merely predict the presence of a crash. 
It further suggests that runtime information may provide larger relative benefits for certain smaller LLMs such as Qwen.

\important{RQ1/2}{The benefits of runtime information are more pronounced when LLMs perform crash diagnosis in addition to predicting its existence, suggesting that runtime information is particularly valuable for reasoning about underlying causes.}

\subsection{RQ2: Runtime info impact on ML libraries and root causes}
\label{subsec: results_rq2}

\autoref{fig: results_CRANE_lines} shows how runtime information affects the LLM performance of crash prediction and diagnosis across ML libraries (used in the crashing line) and root causes. Improvements are measured by the percentage difference in recall between \textit{+RT} and \textit{–RT}.
Recall measures the proportion of actual crashes (i.e., in buggy notebooks) that are correctly predicted and diagnosed.


\autoref{fig: results_CRANE_lines}(a) highlights recall improvements across ML libraries.
Runtime information benefits recall of all LLMs across \emph{PyTorch}, \emph{Scikit-learn}, \emph{Pandas}, and the group of other libraries (i.e., \emph{Seaborn}, \emph{Matplotlib}, \emph{Statsmodels}, \emph{LightGBM}, and \emph{TorchVision}). 
The most notable gains occur for \emph{PyTorch} and \emph{Scikit-learn}, where GPT-5 and Gemini achieve improvements up to 33\pp.
Gemini shows larger recall improvements for \emph{TensorFlow/Keras}, \emph{Scikit-learn}, and \emph{NumPy}, and GPT-5 for \emph{PyTorch}, \emph{Pandas}, and other libraries, aligning with their high overall recall gains in~\autoref{tb: results_overall}(a). 

These findings suggest that while runtime information generally improves crash prediction and diagnosis, its impact varies across both LLMs and ML libraries. This variation likely reflects the differences in each LLM’s internal reasoning strategies and familiarity with specific library semantics, as well as the underlying nature of the crashes. The consistent improvements for \emph{PyTorch}, \emph{Scikit-learn}, and \emph{Pandas} in recall suggest that crashes related to these libraries often depend on runtime-dependent factors that LLMs can effectively use. In contrast, the mixed results for \emph{TensorFlow/Keras} and \emph{NumPy} indicate that \emph{TensorFlow}’s layered abstractions and \emph{NumPy}’s lower-level operations may limit the visibility of relevant runtime information. Finally, the consistent improvements for \emph{other} libraries suggest that, even though the runtime information captures a limited set of data objects from certain libraries, many crashes related to other libraries involve such objects, allowing runtime information to contribute meaningfully.

\important{RQ2/1}{Runtime information improves LLM-based crash prediction and diagnosis across most ML libraries, with the largest recall gains for \emph{PyTorch}, \emph{Scikit-learn}, \emph{Pandas}, and other libraries. The impact varies by LLM and library, reflecting differences in LLM reasoning, familiarity with specific libraries, and crash characteristics.}

\autoref{fig: results_CRANE_lines}(b) shows that runtime information consistently improves recall across all LLMs for crashes caused by \textit{API misuse}, \textit{data confusion}, and \textit{implementation error} (labels provided by JunoBench). This suggests that such crashes depend on the actual runtime state of data or objects, which static code analysis alone cannot fully capture. For instance, shape mismatches of \emph{PyTorch} or \emph{TensorFlow} tensors or incorrect indexing of data frames only manifest when concrete data is loaded, making them inherently dynamic. The observed gains indicate that LLMs can use such runtime information to better identify the presence of these crashes and provide more accurate diagnosis.

In contrast, crashes caused by notebook-specific issues show smaller improvements. This is because many notebook-specific crashes arise from missing cell executions, leading to errors such as undefined variables. Since these errors are easy to predict from the code of the ordered executed cells, additional runtime information offers limited recall benefit.

\input{Main/Sub-content/fig_results_CRANE_libraries_lines}

\important{RQ2/2}{Runtime information helps LLMs predict and diagnose crashes with root causes such as \textit{data confusion}, \textit{API misuse}, and \textit{implementation error}, showcasing its value in identifying crashes that rely on runtime-dependent factors.}

\subsection{Threats to Validity}
\paragraph*{Construct validity}
Our baseline (\emph{without runtime information}) uses ordered executed code cells, which inherently encode runtime context such as execution order. Therefore, improvements from runtime information may be slightly conservative, though this setup reflects a realistic reproducible context. Another threat concerns performance measurements. To mitigate this, we adopted widely used metrics, including precision, recall, F1-score, and accuracy~\cite{yang24mulbugs, li21faulloc, li19deepfl}.

\paragraph*{Internal validity}
LLMs are inherently non-deterministic and may produce different outputs for identical inputs. While we mitigate this by aggregating results from five independent runs, some output variability may remain and could affect the reported performance.
Moreover, CRANE-LLM's effectiveness may vary across LLMs due to their differences in architecture, training data, and alignment. To reduce this threat, we evaluated three different LLMs and observed consistent trends, suggesting that our results are not tied to a specific LLM.
Finally, evaluating diagnosis correctness involves human judgment and may introduce subjectivity. To mitigate this, two experienced evaluators independently assessed outputs and achieved high inter-rater agreement (Cohen’s $\kappa = 0.89$). 

\paragraph*{External validity}
Our experiments use JunoBench, which covers ML notebook crashes related to widely used ML libraries, but may not represent other learning paradigms such as unsupervised or reinforcement learning, domain-specific frameworks, or industrial ML systems.
Since JunoBench consists of publicly available Kaggle notebooks, potential overlap with LLM training data cannot be ruled out.
Moreover, due to the evolving nature of LLM training data and model updates, results with current LLMs may not generalize to future models. However, our focus on relative comparisons and cross-LLM evaluation helps mitigate this concern.

%% file: Main/Sub-content/tb_results_overall.tex
\begin{table*}[t]
    \small
    \centering
    \caption{CRANE-LLM and baseline results on JunoBench without (-RT) and with (+RT) runtime information. Blue/red denotes +RT gains/decreases, with intensity showing change magnitude. Statistical significance using McNemar is reported ($\alpha = 0.05$).}
    \label{tb: results_overall}
    
    \begin{minipage}[t]{\textwidth}
        \centering
        \textbf{(a) Crash prediction and diagnosis (end-to-end) results.}
        \begin{tabular}{c|c|c|c|c|c|c|c|c|c}
            \toprule
            \thead{LLM}  & \thead{Precision\\(-RT)} & \thead{Precision\\(+RT)} & \thead{Recall\\(-RT)} & \thead{Recall\\(+RT)} & \thead{F1\\(-RT)} & \thead{F1\\(+RT)} & \thead{Accuracy\\(-RT)} & \thead{Accuracy\\(+RT)} & \thead{McNemar \\ (p-value)}\\ 
            \midrule
            \makecell{Gemini} 
            & 57.5\% & \comptext{0.575}{0.633}
            & 62.2\% & \comptext{0.622}{0.730}
            & 59.7\% & \comptext{0.597}{0.678}
            & 58.1\% & \comptext{0.581}{0.653}
            & \textcolor{teal}{0.0335} (\greenarrow Significant)\\
            \hline
        
            \makecell{Qwen} 
            & 67.5\% & \comptext{0.675}{0.838}
            & 50.5\% & \comptext{0.505}{0.559}
            & 57.7\% & \comptext{0.577}{0.670}
            & 63.1\% & \comptext{0.631}{0.725}
            & \textcolor{teal}{0.0016} (\greenarrow Significant)\\
            \hline
        
            \makecell{GPT-5}  
            & 75.0\% & \comptext{0.750}{0.798}
            & 56.8\% & \comptext{0.568}{0.712}
            & 64.6\% & \comptext{0.646}{0.752}
            & 68.9\% & \comptext{0.689}{0.766}
            & \textcolor{teal}{0.0191} (\greenarrow Significant)\\
            \hline
        \end{tabular}
    \end{minipage}

    \begin{minipage}[t]{\textwidth}
        \centering
        \textbf{\\(b) Crash prediction-only (crash existence prediction) results.\\}
        \begin{tabular}{c|c|c|c|c|c|c|c|c|c}
            \toprule
            \thead{LLM}  & \thead{Precision\\(-RT)} & \thead{Precision\\(+RT)} & \thead{Recall\\(-RT)} & \thead{Recall\\(+RT)} & \thead{F1\\(-RT)} & \thead{F1\\(+RT)} & \thead{Accuracy\\(-RT)} & \thead{Accuracy\\(+RT)} & \thead{McNemar \\ (p-value)}\\ 
            \midrule
            \makecell{Gemini} 
            & 63.8\% & \comptext{0.638}{0.678}
            & 81.1\% & \comptext{0.811}{0.892}
            & 71.4\% & \comptext{0.714}{0.770}
            & 67.6\% & \comptext{0.676}{0.734}
            & {0.0659} (Not Significant)\\
            \hline
        
            \makecell{Qwen} 
            & 72.4\% & \comptext{0.724}{0.860}
            & 64.0\% & \comptext{0.640}{0.667}
            & 67.9\% & \comptext{0.679}{0.751}
            & 69.8\% & \comptext{0.698}{0.779}
            & \textcolor{teal}{0.0113} (\greenarrow Significant)\\
            \hline
        
            \makecell{GPT-5}  
            & 79.0\% & \comptext{0.790}{0.815}
            & 71.2\% & \comptext{0.712}{0.793}
            & 74.9\% & \comptext{0.749}{0.804}
            & 76.1\% & \comptext{0.761}{0.806}
            & {0.1214} (Not Significant)\\
            \hline
        \end{tabular}
    \end{minipage}
    
        
        
\end{table*}

%% file: Main/Sub-content/tb_results_diagnosis_subcat.tex
\begin{table}[t]
    \small
    \centering
    \caption{Breakdown of incorrect diagnosis predictions (\emph{Total}) into \emph{Partially Correct (PC)} and \emph{Reasoning Wrong (RW)}.}
    \label{tb: results_diagnosis_subcat}
    \begin{tabular}{c|c|c|c|c|c|c}
        \toprule
        \thead{LLM}  & \thead{Total\\(-RT)} & \thead{Total\\(+RT)} & \thead{PC\\(-RT)} & \thead{PC\\(+RT)} & \thead{RW\\(-RT)} & \thead{RW\\(+RT)}\\ 
        \midrule
        \makecell{Gemini} 
        & 21 & \textcolor{red!100}{18}
        & 1 & \textcolor{red!100}{0}
        & 20 & \textcolor{red!100}{18}\\
        \hline
    
        \makecell{Qwen} 
        & 15 & \textcolor{red!100}{12}
        & 1 & {1}
        & 14 & \textcolor{red!100}{11}\\
        \hline
    
        \makecell{GPT-5}  
        & 16 & \textcolor{red!100}{10}
        & 5 & \textcolor{red!100}{3}
        & 11 & \textcolor{red!100}{7}\\
        \hline
    \end{tabular}
\end{table}

%% file: Main/Sub-content/fig_results_CRANE_libraries_lines.tex
\begin{figure}[t]
    \centering
    \small
    \begin{minipage}{\linewidth}
        \centering
        \includegraphics[width=\linewidth]{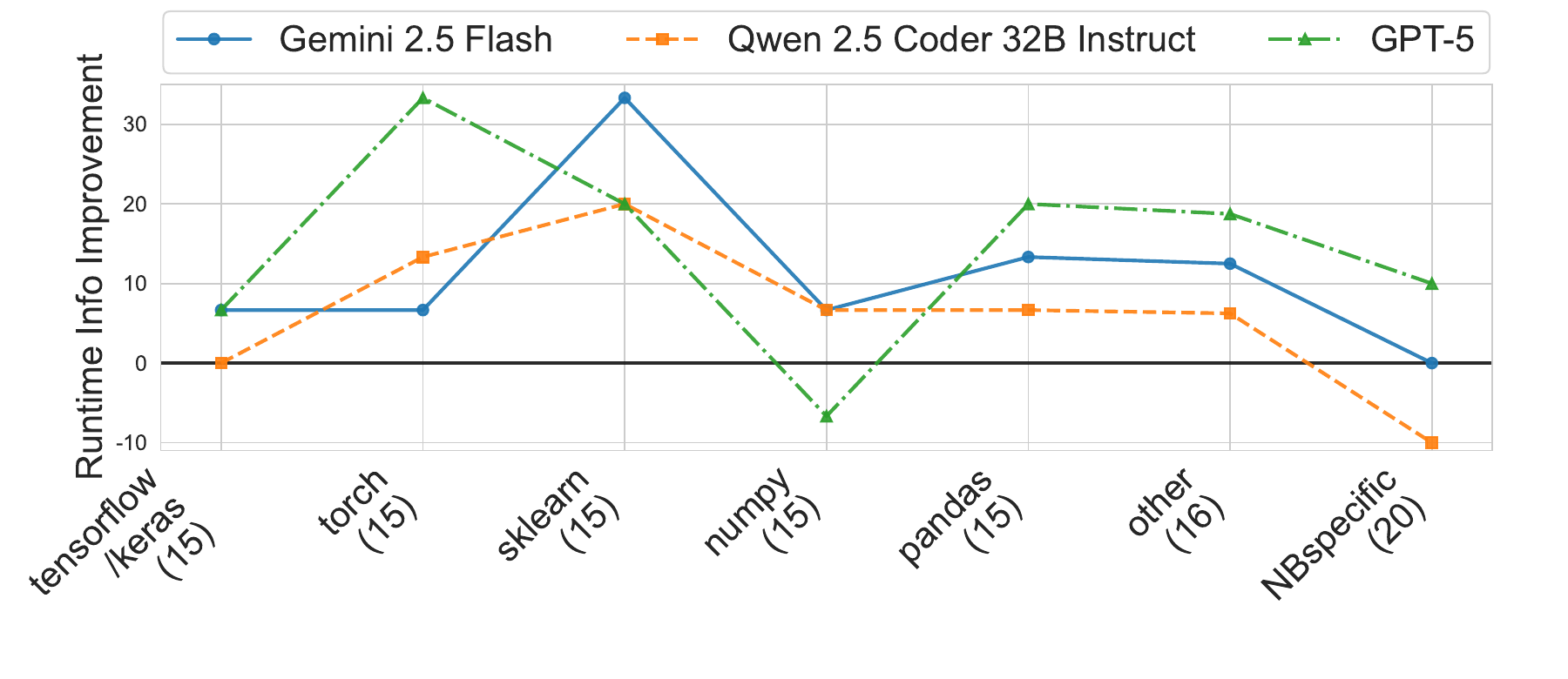} 
        \vspace{-0.8cm}
        {\\(a) By libraries, \emph{other} include \emph{Seaborn} (6), \emph{Matplotlib} (6), \emph{Statsmodels} (2), \emph{LightGBM} (1), and \emph{TorchVision} (1)\\}
    \end{minipage}
    \vspace{0.5cm}
    \begin{minipage}{\linewidth}
        \centering
        \includegraphics[width=0.6\linewidth]{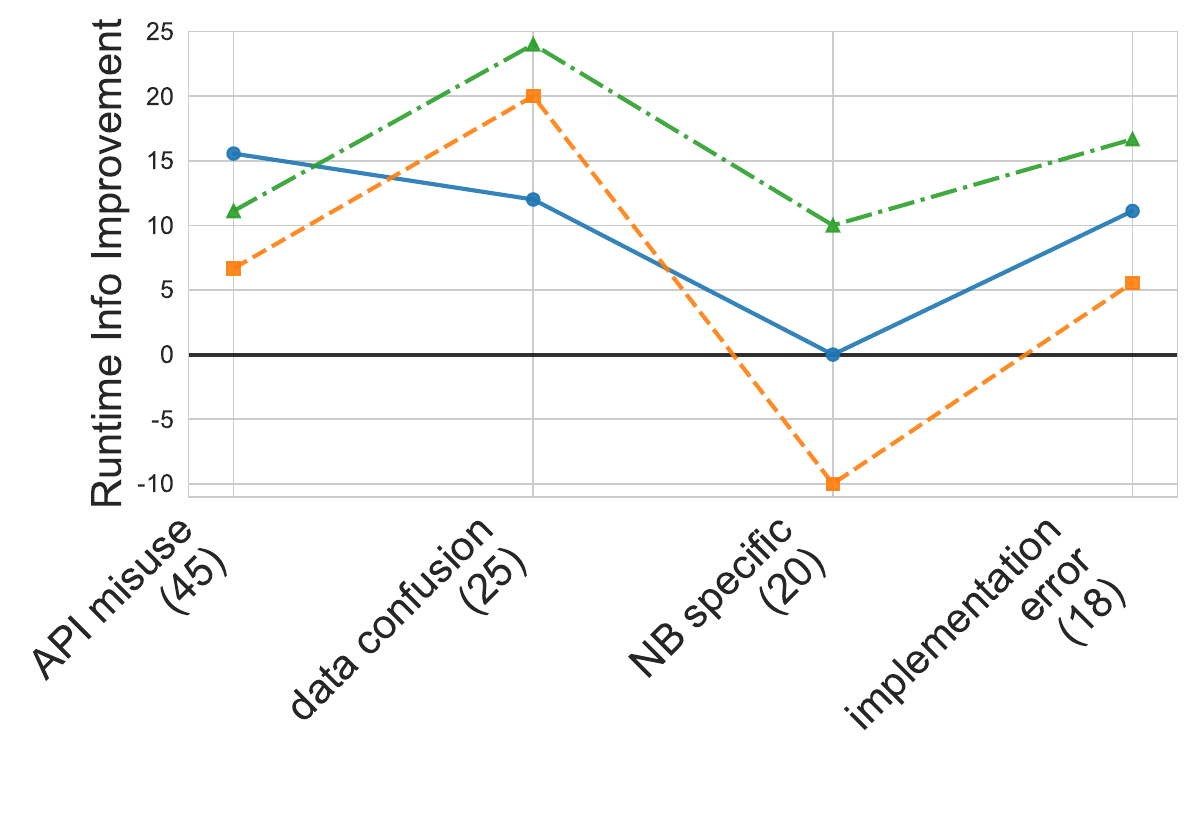} 
        \vspace{-0.5cm}
        {\\(b) By root causes\\}
    \end{minipage}
    \vspace{-0.8cm}
    \caption{Recall improvements of the LLM (+RT) on JunoBench.}
    \label{fig: results_CRANE_lines}
\end{figure}

%% file: Main/Content/6_discussion.tex
\section{Discussion}
\label{Sec: Discussion}
This section analyzes several JunoBench notebooks from our experiments, examines early crash detection enabled by CRANE-LLM, and studies the impact of runtime information categories and additional API information on CRANE-LLM.

\subsection{Runtime info helps LLMs for crash prediction and diagnosis}
\autoref{subsec: results_rq1} shows that runtime information significantly improves CRANE-LLM's crash prediction and diagnosis performance across all evaluated LLMs.
For example, in the JunoBench notebook \lstinline[style=mystylecode]{sklearn_6_reproduced.ipynb}, the target cell crashes when a fitted \lstinline[style=mystylecode]{RandomForestRegressor} model is used to predict on a test dataset with mismatched feature dimensions.
Using only source code, the feature dimensions of the training and test datasets are unavailable because both are loaded from external files, leading GPT-5 to incorrectly predict no crash. 
In contrast, runtime information reveals that the random forest model is trained with 58 features (\lstinline[style=mystylecode]{'n_features_in_': 58}) while the test data contains 53 columns (\lstinline[style=mystylecode]{'shape': (1459, 53)}),
enabling correct crash identification by GPT-5. 
This example demonstrates that runtime information provides essential data attributes (e.g., data shapes and model properties) that cannot be inferred from static code.
It also shows that LLMs (e.g., GPT-5) encode useful ML library semantics (e.g., feature consistency constraints) without requiring explicit rules. 
This implicit domain knowledge, combined with runtime information, enables LLMs to accurately predict and diagnose crashes.

\subsection{Runtime info enables cross-library prediction and diagnosis}
\autoref{subsec: results_rq2} shows that runtime information extracted from common data structures (e.g., \emph{NumPy} arrays, \emph{Pandas} dataframes) and ML-specific objects (e.g., \emph{TensorFlow/Keras}, \emph{PyTorch}, and \emph{Scikit-learn}) can improve CRANE-LLM's crash prediction and diagnosis not only for these libraries but also for others such as \emph{Seaborn} and \emph{Matplotlib}, which are not directly summarized (see~\autoref{sec:runtime-info-extract}).
For example, in \lstinline[style=mystylecode]{seaborn_6_reproduced.ipynb}, a \lstinline[style=mystylecode]{sns.violinplot(df, x='species')} call crashes because the \lstinline[style=mystylecode]{'species'} column is categorical instead of numeric. 
Without runtime information, Gemini and Qwen failed to predict the crash because \lstinline[style=mystylecode]{df} originates from an external dataset. 
With runtime information showing that the \lstinline[style=mystylecode]{'species'} contains three unique categorical values, both LLMs correctly predicted the crash and diagnosed the crash as being caused by passing the categorical column as the \lstinline[style=mystylecode]{x} argument to \lstinline[style=mystylecode]{sns.violinplot} without providing a numeric \lstinline[style=mystylecode]{y} argument, even though the API requires at least one numeric variable.
This indicates that runtime properties of shared data structures enable effective crash prediction even for library APIs not explicitly covered in the runtime extractions, and demonstrating the broader generality of our approach.

\subsection{CRANE-LLM's early crash detection (crash prediction) saves execution time by avoiding kernel state corruption}
CRANE-LLM introduces a small inference overhead (e.g., avg 1.6s per GPT-5 query), but this must be weighed against the extra execution time it can help avoid. In interactive notebook environments, crashes often corrupt the kernel state, requiring developers to rerun \emph{all} previously executed cells (see~\autoref{sec_background: kernel_corruption}). 
Across 111 buggy JunoBench notebooks, comparing cumulative execution time of all cells preceding the crashing cell with GPT-5 query latency yields a net saving of 1,103s (approximately 18 minutes). While this comparison is not intended to precisely measure end-to-end developer productivity, it provides a conservative, observable estimate.

Another time saving comes from preventing execution of the crashing cell itself. 
For example, in notebook \lstinline[style=mystylecode]{torchvision_1_reproduced.ipynb}, the target cell builds, trains, and evaluates a \lstinline[style=mystylecode]{torchvision.models.resnet18} network. The crash occurs during evaluation (after prolonged training) due to mismatched input shape (\lstinline[style=mystylecode]{[N, 3, 256, 256, 1]}, revealed by runtime information) with model requirements of 4-dimensional shape (\lstinline[style=mystylecode]{[N, C, H, W]}).
While executing this cell takes 82 seconds for a single epoch before the crash occurs, CRANE-LLM predicts the issue within two seconds. 

Overall, these preliminary measurements are conservative lower bounds, as they exclude kernel restart time and underestimate cell execution time due to benchmark-level reductions in input data sizes and training epochs. The results nonetheless show that CRANE-LLM’s inference cost is small compared to the execution time it can prevent, motivating further evaluation in real development settings.

\subsection{Runtime info categories and API documentation}
We further analyzed the contribution of runtime information categories, which we grouped into: (1) structural information (e.g., tensor or dataset shapes, sizes, and counts), (2) type-level information (e.g., object types, data types, dtypes, and schema properties), and (3) value-semantic information (e.g., summarized statistics or concrete runtime values). This categorization is motivated by a prior empirical study of ML notebook crashes~\cite{wang25mlcrash}. To study their impact, we performed a small ablation study in which each category was removed from CRANE-LLM and evaluated on the crash prediction-only task, since the diagnosis evaluation requires substantial manual effort.
The ablation results show that Qwen exhibited performance degradation when any runtime category was removed, with the largest drops observed for structural and type-level information. In contrast, GPT-5 benefited most from value-semantic runtime information. These findings suggest that different LLMs rely on different forms of runtime evidence when reasoning about notebook crashes. More broadly, the results indicate that runtime-aware debugging systems may benefit from adapting the representation and prioritization of runtime context to the target LLM rather than treating all runtime information as equally useful.

Motivated by prior work on LLM hallucination in code understanding~\cite{zhang2025sirenllmhallu, zhang2025llmhallu, liu2024llmhallu} and API documentation grounding~\cite{jain2024mitigatingcodellmhallucinations, li2025empiricalapidoc}, we additionally evaluated whether augmenting CRANE-LLM with API information improves its crash prediction-only performance. Specifically, we extracted structured documentation for relevant API calls appearing in the target cell and incorporated this information into the LLM input.
Contrary to expectations, API documentation grounding did \emph{not} improve performance for any evaluated LLM, while substantially increasing the input context size (73.6\% additional tokens on average) and therefore inference cost. One possible explanation is that runtime information already provides a sufficiently concrete execution context for crash existence prediction. Additional documentation may introduce redundant or weakly relevant information that interferes with effective LLM reasoning. More broadly, these results suggest that adding auxiliary contextual information does not necessarily improve LLM-based bug reasoning and should be evaluated carefully against its associated cost. 
More detailed results on the ablation study and API documentation grounding experiment can be found in our GitHub repository~\cite{wang2025code}.

%% file: Main/Content/7_relatedwork.tex
\section{Related Work}
\paragraph{Bug prediction/detection for ML scripts}
Predicting or detecting bugs in ML programs is challenging due to implicit data constraints, complex ML library APIs, and opaque ML model behaviors.
Several works have studied bug prediction or detection in ML programs written in Python scripts, particularly for libraries such as \emph{TensorFlow/Keras} and \emph{PyTorch}. 

\emph{Static analysis approaches} predict bugs before execution by reasoning over source code and symbolic properties. 
Examples include ShapeTracer~\cite{liu22shapetracer}, PyTea~\cite{jhooPyteaStaticAnalyzer2022}, work by Baker et al.~\cite{bakerDetectFixVerify2022}, NeuraLint~\cite{Nikanjam21graph}, Theia~\cite{manke25structurebugs}, and PYSIASSIST~\cite{hongInvestigatingDetectingSilent2024}.
While effective for known bug patterns, static methods struggle with failures that depend on concrete runtime states.
\emph{Dynamic approaches} analyze execution behavior to detect runtime issues such as shape mismatches~\cite{vermaShapeFlowDynamicShape2020}, and faults during model training~\cite{Wardat21deeplocalize, Braiek23fnn, zhang21autotrainer, Wardat22deepdiagnosis}.
However, they typically focus on specific bug categories and identify faults only during or after execution.
\emph{Learning-based techniques} train classifiers with program features or execution context to predict ML-specific bug patterns~\cite{wuTensfaDetectingRepairing2021, cao22deepFD, jahan2025improveddetectiondiagnosisfaults, sharmin25sa}.
These methods depend on curated training datasets and predefined fault types, limiting generalization to unseen bugs.
More recently, \emph{LLM-based approaches} have explored prompt-based bug prediction for ML programs~\cite{weiDemystifyingDetectingMisuses2024, caoStudyPromptDesign2023}.
Although LLMs can reason about ML code without handcrafted rules, existing approaches rely only on source code and do not employ execution context, making them less effective for runtime-dependent failures.

\paragraph{Bug prediction/detection for notebooks}
ML notebooks are different from script-based ML programs due to their interactive execution model and persistent kernel state. 
Several studies apply \emph{static analysis} to predict data leakage-related issues using data-flow analysis and abstract interpretation~\cite{suboticStaticAnalysisFramework2022, suboticStaticallyDetectingData2022, yang23leakagenb, Drobnjakovic24leakagenb}, or API misuses with rule-based approaches~\cite{liblit23left}. 
Other works use \emph{execution monitoring} with static analysis for general notebook analysis rather than ML-specific issues, such as unsafe cell orderings~\cite{macke21finegrain} and regression bugs~\cite{yao2025regressiontestingframeworkautomated}.
Recently, \emph{LLM-based} coding assistants have been proposed for interactive notebook debugging after failures occur~\cite{grotov2024debugsmarterharderai, levin25chatdbg}. 
These systems (e.g., ChatDBG~\cite{levin25chatdbg}) operate in a post-hoc setting where users iteratively diagnose errors through conversational interaction and re-execution. 
In contrast, our work focuses on proactive crash prediction before executing a target cell, aiming to prevent failures rather than debug them after occurrence.


Prior work~\cite{wang24runtime} first investigated augmenting LLMs with runtime information for ML bug prediction and reported encouraging results on \emph{TensorFlow} shape-related errors. However, it was limited to a single library and a specific bug type. 
Their evaluation dataset was built from standalone \texttt{.py} files converted into notebook cells, which may not reflect crashes in real-world ML notebooks. 
In contrast, our work systematically evaluates runtime augmentation on a broader set of real ML notebook crashes spanning multiple libraries and crash types.

\paragraph{Summary and our contribution}
Prior work on ML bug prediction or detection largely relies on static analysis, pre-defined rules, hand-crafted verification conditions, or supervised learning over curated training dataset and fixed fault categories.
While LLM-based methods show promise, they have been applied primarily in a static setting or as interactive debugging assistants.
CRANE-LLM instead targets crash prediction and diagnosis for ML notebooks. 
It augments LLMs with source code and structured runtime information extracted from the notebook kernel to predict crashes before cell execution and generate natural-language diagnoses.

%% file: Main/Content/8_conclusion.tex
\section{Conclusions}

We present CRANE-LLM, a runtime-augmented LLM-based approach for crash prediction and diagnosis in ML notebooks.
CRANE-LLM integrates runtime information extracted from the notebook kernel, such as object types and tensor shapes, together with static code context into LLM prompts, enabling the LLM to predict if a target cell will crash before execution and provide explanatory diagnoses.
Our evaluation on JunoBench shows that incorporating runtime information significantly improves crash prediction and diagnosis across multiple state-of-the-art LLMs compared to using static code alone.
The improvements are more pronounced when diagnosis is required, indicating that runtime context is particularly valuable for reasoning about crash causes than merely predicting the presence of a crash.
We also observe that the impact of runtime information varies across ML libraries, root causes, and LLMs, 
suggesting that its benefits depend on both the nature of the crash and the LLM's reasoning characteristics.